\newcommand{\logg}{$\log\,g$}
\newcommand{\teff}{$T_{\rm eff}$}

\documentclass{aa}
\usepackage{txfonts}
\usepackage{graphicx}
\begin{document}

\title{
New \ion{Fe}{ii} energy levels from stellar spectra
}

\author{
F.\, Castelli
\inst{1}
\and
R.L.\, Kurucz
\inst{2}
}

\offprints{F. Castelli}

\institute{
Istituto Nazionale di Astrofisica--
Osservatorio Astronomico di Trieste, Via Tiepolo 11,
I-34131 Trieste, Italy\\
\email{castelli@oats.inaf.it}
\and
Harvard-Smithsonian Center for Astrophysics, 60 Garden Street,
Cambridge, MA 02138, USA
}

\date{}

\abstract
{}
{The spectra of B-type and early A-type stars show  numerous
unidentified lines in the whole optical range, especially in the 
5100-5400\,\AA\ interval.  Because \ion{Fe}{ii} transitions to
high energy levels should be observed in this region, we used semiempirical
predicted wavelengths and gf-values of \ion{Fe}{ii} to identify  unknown lines.   
}
{Semiempirical line data for \ion{Fe}{ii} computed by Kurucz are used
to synthesize the spectrum of the slow-rotating, \ion{Fe}-overabundant
CP star HR\,6000.
}
{
We determined a total of 109  new 4f levels for \ion{Fe}{ii} 
with energies ranging from 122\,324\,cm$^{-1}$ to 128\,110\,cm$^{-1}$. They belong
to the \ion{Fe}{ii} subconfigurations 3d$^{6}$($^{3}$P)4f (10 levels), 3d$^{6}$($^{3}$H)4f (36\,levels), 
3d$^{6}$($^{3}$F)4f
(37\,levels), and 3d$^{6}$($^{3}$G)4f (26\,levels). 
We also found 14 even levels from 4d (3\,levels), 5d (7\,levels), and 6d (4\,levels)
configurations. The new levels 
have allowed us to identify more than the 50\,\% of 
previously unidentified lines of HR\,6000 in the 
wavelength region 3800-8000\,\AA. Tables listing the new energy
levels are given in the paper; tables listing the spectral lines
with $\log\,gf$\,$\ge$\,$-$1.5 that are transitions to the 4f energy levels are given
 in the Online Material. These new
levels produce 18000 lines throughout the spectrum from the
ultraviolet to the infrared.
}
{}


\maketitle{}

\section{Introduction}

In a previous paper (Castelli, Kurucz \& Hubrig, 2009) (Paper I) we have
determined 21 new 3d$^{6}$($^{3}$H)4f high energy levels of \ion{Fe}{ii} on the basis
of predicted energy levels, computed $\log\,gf$ values for \ion{Fe}{ii}, and 
unidentified lines in UVES high resolution, high signal-to-noise spectra of 
HR\,6000 and 46\,Aql. Both stars are iron overabundant CP stars  and have  
rotational velocity v{\it sini} of the order of 1.5\,km\,s$^{-1}$ and 
1.0\,km\,s$^{-1}$, respectively.

In this paper we continue the effort to determine
new high-energy levels of \ion{Fe}{ii}.
We used the same spectra and models for HR\,6000 that we 
adopted in Paper\,I,
together with \ion{Fe}{ii} line lists which include
transitions between observed-observed, observed-predicted, and 
predicted-predicted  energy levels. 
In this paper we increase the number of the new energy levels from 
the 21  listed in Paper\,I, to a total of 109 energy levels, 
which belong to the \ion{Fe}{ii} subconfigutations:
3d$^{6}$($^{3}$P)4f (10 levels), 3d$^{6}$($^{3}$H)4f (36 levels), 
3d$^{6}$($^{3}$F)4f
(37 levels), and 3d$^{6}$($^{3}$G)4f (26 levels), and 14 levels
from the even configurations 4d (3\,levels), 5d (7\,levels), and
6d (4\,levels).  The new levels 
have allowed us to identify more than the 50\,\% of the 
previously unidentified lines in the 
wavelength region 3800-8000\,\AA\  of HR\,6000 (Castelli \& Hubrig, 2007).  
The method that we adopted  to determine the new energy levels is the
same as described in Paper\,I. It is recalled here in Sect.\,3.
The  comparison of the observed spectrum of HR\,6000 with the
synthetic spectrum which includes the new \ion{Fe}{ii} 
lines  is available on the Castelli web site\footnote{http://wwwuser.oat.ts.astro.it/castelli/hr6000new/hr6000.html}.

\section{The star HR\,6000}

According to Paper\,I, the CP star HR\,6000 (HD\,144667) has 
an estimated
rotational velocity of 1.5\,km\,sec$^{-1}$.  The model stellar
parameters  for an individual 
abundance ATLAS12 (Kurucz 2005) model are \teff=13450\,K, \logg=4.3.
In addition to the large iron overabundance [$+$0.9],
overabundances of \ion{Xe} ([$+$4.6]), \ion{P} ($>$[$+$1.5]),
\ion{Ti} ([$+$0.55]), \ion{Cr} ([$+$0.2]), \ion{Mn} ([$+$1.5]), 
Y ([$+$1.2]), and Hg ([$+$2.7]) were observed.
This peculiar chemical composition, together with the
underabundances of He, C, N, O, Al, Mg, Si, S, Cl, Sc, V,
Co, Ni, and Sr gives rise to an optical line spectrum 
very rich in \ion{Fe}{ii} lines, with transitions involving
upper energy levels close to the ionization limit (Johansson 2009). 
Also numerous \ion{Fe}{i}
and \ion{Fe}{iii} lines are observable in the spectrum.

\section{The method}

To determine the new energy levels we used  high-resolution 
 UVES spectra of HR\,6000 (see Paper\,I), 
the corresponding synthetic spectrum,  and the list 
of the computed transitions with predicted values for levels with  
no experimentally available energies. 
Predicted energy levels and  $\log\,gf$ values
were computed by Kurucz with his version of the Cowan (1981) code
(Kurucz 2009). The calculation included 46 even configurations
d$^{7}$, d$^{6}$4s$-$9s, d$^{6}$4d$-$9d, d$^{6}$5g$-$9g, 
d$^{6}$7i-9i, d$^{6}$9l, d$^{5}$4s$^{2}$, d$^{5}$4s5s$-$9s,
d$^{5}$4s4d$-$9d, d$^{5}$4s5g-9g, d$^{5}$4s7i$-$9i, 
d$^{5}$4s9l, d$^{4}$4s$^{2}$4d, and d$^{5}$4p$^{2}$ with 19771
levels least-squares fitted to  418 known levels.
The 39 odd configurations included d$^{6}$4p$-$9p, d$^{6}$4f$-$9f,
d$^{6}$6h$-$9h, d$^{6}$8k$-$9k, d$^{5}$4s4p$-$9p, d$^{5}$4s4f$-$9f,
d$^{5}$4s6h$-$9h, d$^{5}$4s8k-9k, d$^{4}$4s$^{2}$4p$-$5p, and
d$^{4}$4s$^{2}$4f with 19652 levels least-squares fitted to
596 known levels. The calculations were done in LS coupling with
all configuration interactions included, with scaled Hartree-Fock
starting guesses, and with Hartree-Fock transition integrals.
A total of 7080169 lines were saved from the transition array
of which 102833 lines are between known levels and have good
wavelengths.
The computed line list was sorted into tables of all the strong
lines connected to every predicted level.
When a given predicted  level gives rise to at least two
\ion{Fe}{ii} lines having $\log\,gf$\,$\ge$\,$-$1.0, we selected 
one of these transitions and searched
in the spectrum for those unidentified lines which have wavelength
within $\pm$50\,\AA\ and residual flux
within about $\pm$ 5\% of those of the selected predicted line.  
From the observed wavelength of one of these unidentified lines and 
from the known energy of the lower or upper level of the predicted transition, 
we derived a possible energy 
for the predicted level. If most of transitions obtained with this energy correspond to
lines observed in the spectrum,  we kept the tentative energy value as 
a real value, otherwise we repeated the procedure using another line taken from the 
unidentified  ones, and continued the searching  until we found that energy
for which most of the predicted lines correspond to the observed lines.
Whenever one or more new levels were found, the whole semiempirical
calculation was repeated to produce improved predicted wavelengths
and $\log\,gf$-values. Because all configuration interactions were
included, and because the mixing is exceptionally strong in the 4d and 5d
configurations,
every new level changed the predictions. Mixing between close levels
can produce large uncertainties in the $\log\,gf$ values for lines
that involve those levels.

This procedure is very successfull for levels which produce two or
more transitions with $\log\,gf$\,$>$\,0.0, but becomes more and more
difficult as the intensity of the predicted lines decreases. 
In fact, weak lines are usually blended with stronger components,
so that the method may fail in these cases.

\section{The new energy levels}

The new energy levels of the  3d$^{6}$($^{3}$P)4f, 3d$^{6}$($^{3}$H)4f,
3d$^{6}$($^{3}$F)4f, and 3d$^{6}$($^{3}$G)4f subconfigurations 
and from the even configurations 3d$^{6}$4d, 3d$^{6}$5d, and 3d$^{6}$6d are listed
in Tables\,1$-$5. 
 Because the 3d$^{6}$4f states of \ion{Fe}{ii} tend to appear in pairs we have used
the j$_{c}$[K]$_{j}$ notation of  jK coupling for them,
where {\bf j}$_{c}$ is the total angular momentum of the core and 
{\bf K}={\bf J$_{c}$}$+${\bf l} is the coupling of {\bf J$_{c}$} with
the orbital angular momentum {\bf l} of the active electron. The level
pairs correspond to the two separate values of the total angular
momentum {\bf J} obtained when the spin s=$\pm$1/2 of the active
electron is added to {\bf K}.
 The positive energies are those obtained
by comparing observed and predicted line profiles, as described in Sect.\,3
and shown in Fig.\,2. 
The energies between parentheses in Tables\,1$-$4 are  predicted  values for which we have been not able
to find the corresponding observed level. The reason for the failure is
that either all the lines from the energy level are weak or, even if some of the 
transitions are predicted as 
moderately strong ($\log\,gf$\,$>$\,0.0), they are blended with 
other stronger components, so that their identification is uncertain. 
The columns with label ``c$-$o'' in Tables 1-5 show the difference between the
predicted and observed energy levels.  

The 4d even energy levels listed in Table\,5 give rise to some of the transitions
listed in the Online Material.
The strongest transitions related with the  5d, and 6d even energy  levels 
occur in the  6000-8000\,\AA\, region 
and in the 4000-5000\,\AA\, region, respectively.
The transitions to the odd energy levels are discussed in Sect.\,5

The observed energy levels, the least squares fits, the predicted energy 
levels, and the line lists can be found on the Kurucz 
web site\footnote{http://kurucz.harvard.edu/atoms/2601}. The observed
levels come from the following sources: Johansson (1978), Sugar \& Corliss (1985),
Adam et al. (1987), Johansson \& Baschek (1988), Johansson (1988, private
communication), Rosberg \& Johansson (1992), Castelli, Johansson \& Hubrig (2008),
Castelli, Kurucz, Hubrig (2009), and this work. The calculations
on the web site are updated whenever there are improvements to the
energy levels.

\begin{table*}{}
\begin{flushleft}
\caption {\ion{Fe}{ii} energy levels for the  3d$^{6}$\,($^{3}$P)4f subconfiguration. Energies
between parentheses are predicted values.} 
\begin{tabular}{crcrcrcrcrcrrrr}
\noalign{\smallskip}\hline
\multicolumn{1}{c}{Design-}&
\multicolumn{1}{c}{J}&
\multicolumn{1}{c}{Energy}&
\multicolumn{1}{c}{c$-$o}&
\multicolumn{1}{c}{Design-}&
\multicolumn{1}{c}{J}&
\multicolumn{1}{c}{Energy}&
\multicolumn{1}{c}{c$-$o}&
\multicolumn{1}{c}{Design-}&
\multicolumn{1}{c}{J}&
\multicolumn{1}{c}{Energy}&
\multicolumn{1}{c}{c$-$o}
\\
\multicolumn{1}{c}{ation}&
\multicolumn{1}{c}{}&
\multicolumn{1}{c}{cm$^{-1}$}&
\multicolumn{1}{c}{cm$^{-1}$}&
\multicolumn{1}{c}{ation}&
\multicolumn{1}{c}{}&
\multicolumn{1}{c}{cm$^{-1}$}&
\multicolumn{1}{c}{cm$^{-1}$}&
\multicolumn{1}{c}{ation}&
\multicolumn{1}{c}{}&
\multicolumn{1}{c}{cm$^{-1}$}&
\multicolumn{1}{c}{cm$^{-1}$}
\\
\noalign{\smallskip}\hline
2[5]   &11/2  & 122351.810 &$-$20.236 \\
       &9/2   & 122324.142 &$-$18.980 \\
\\
2[4]   &9/2   &122355.116 &$-$6.685   & 1[4]&  9/2 &123629.520&$-$4.606\\
       &7/2   &122355.553 &$-$6.801   &     &  7/2 &123637.833&$-$6.417\\
\\
2[3]   &7/2   & 122351.488 &$-$18.489  & 1[3]&  7/2 &123615.875 &$-$2.642 & 0[3] & 7/2 &(124167.229) &\\
       &5/2   & (122353.541) & &    &  5/2 &123649.493 &$-$5.687 &      & 5/2 & 124157.060&$+$15.841\\
\\
2[2]   &5/2   &(122342.921) &   & 1[2] & 5/2 &(123637.063) &\\
       &3/2   &(122336.098) &   &      & 3/2 &(123646.360) &\\
\\
2[1]   & 3/2 &(122358.405)  & \\
       & 1/2 &(122332.608)  &\\
\hline
\noalign{\smallskip}
\end{tabular}
\end{flushleft}
\end{table*}

\begin{table*}{}
\begin{flushleft}
\caption {\ion{Fe}{ii} energy levels for the 3d$^{6}$\,($^{3}$H)4f subconfiguration. Energies
between parentheses are predicted values.} 
\begin{tabular}{crcrcrcrcrcrrrr}
\noalign{\smallskip}\hline
\multicolumn{1}{c}{Design-}&
\multicolumn{1}{c}{J}&
\multicolumn{1}{c}{Energy}&
\multicolumn{1}{c}{c$-$o}&
\multicolumn{1}{c}{Design-}&
\multicolumn{1}{c}{J}&
\multicolumn{1}{c}{Energy}&
\multicolumn{1}{c}{c$-$o}&
\multicolumn{1}{c}{Design-}&
\multicolumn{1}{c}{J}&
\multicolumn{1}{c}{Energy}&
\multicolumn{1}{c}{c$-$o}
\\
\multicolumn{1}{c}{ation}&
\multicolumn{1}{c}{}&
\multicolumn{1}{c}{cm$^{-1}$}&
\multicolumn{1}{c}{cm$^{-1}$}&
\multicolumn{1}{c}{ation}&
\multicolumn{1}{c}{}&
\multicolumn{1}{c}{cm$^{-1}$}&
\multicolumn{1}{c}{cm$^{-1}$}&
\multicolumn{1}{c}{ation}&
\multicolumn{1}{c}{}&
\multicolumn{1}{c}{cm$^{-1}$}&
\multicolumn{1}{c}{cm$^{-1}$}
\\
\noalign{\smallskip}\hline
6[9]   & 19/2 & 122954.180&$+$14.465\\
       & 17/2 & 122952.730&$+$20.251\\
\\
6[8]   & 17/2 & 123007.910 &$+$26.752 &  5[8] & 17/2 & 123219.200&$-$10.198 \\
       & 15/2 & 122910.920 &$-$16.531 &       & 15/2 & 123193.090&$-$17.864 \\
\\
6[7]   & 15/2 & 123018.430 &$+$34.439 &  5[7]& 15/2 & 123238.440&$-$6.653   &4[7] & 15/2 & 123396.250&$-$33.027\\
       & 13/2 & 123015.400 &$+$40.333 &      & 13/2 & 123168.680&$-$33.645   &     & 13/2 & 123355.490&$-$36.436\\
\\
6[6]   &13/2  & 122990.620 &$-$2.720 &  5[6]& 13/2 & 123249.650&$-$6.519   &4[6] &13/2 & 123414.730&$-$32.244\\
       &11/2  & 123037.430 &$+$26.878 &      & 11/2 & 123270.340&$+$0.899   &     &11/2 & 123427.119&$-$33.418\\
\\
6[5]   &11/2  & 123002.288 &$+$33.455 &  5[5]& 11/2 & 123251.470&$-$1.320   &4[5] & 11/2 & 123441.100&$-$26.889\\
       &9/2   & 123026.350 &$+$18.587 &      &  9/2 & 123269.378&$+$2.937   &     &9/2  & 123435.468&$-$17.705\\
\\
6[4]   &9/2   & 122988.215 &$+$30.836 &  5[4]&  9/2 & 123258.994&$-$1.556   &  4[4] & 9/2 & 123460.690&$-$26.898\\
       &7/2   & 122980.408 &$+$26.752 &     &  7/2  & 123258.021&$-$1.362   &       &7/2 & 123435.277&$-$16.103\\
\\
6[3]   &7/2   & 122946.419 &$+$21.403 &   5[3] & 7/2 & 123235.165&$+$3.471   & 4[3] & 7/2 & 123451.449&$-$21.115 \\
       &5/2   &(122967.896)&          &        & 5/2 & (123248.017)&  &     & 5/2 &  123430.181&$-$16.906\\
\\
       &      &           & &   5[2] & 5/2 & 123211.159&$-$1.017 &   4[2] & 5/2 & (123401.927) \\
       &      &           & &        & 3/2 & 123213.323&$-$12.585 &        & 3/2 & (123384.857)&\\
\\
       &      &           & &        &    &            &     &4[1] & 3/2& (123356.410)    \\
       &      &           & &        &    &            &     &  & 1/2 &   (123343.705)    \\
\hline
\noalign{\smallskip}
\end{tabular}
\end{flushleft}
\end{table*}

\begin{table*}{}
\begin{flushleft}
\caption {\ion{Fe}{ii} energy levels for the  3d$^{6}$\,($^{3}$F)4f subconfiguration. Energies
between parentheses are predicted values.}
\begin{tabular}{crcrcrcrcrcrrrr}
\noalign{\smallskip}\hline
\multicolumn{1}{c}{Design-}&
\multicolumn{1}{c}{J}&
\multicolumn{1}{c}{Energy}&
\multicolumn{1}{c}{c$-$o}&
\multicolumn{1}{c}{Design-}&
\multicolumn{1}{c}{J}&
\multicolumn{1}{c}{Energy}&
\multicolumn{1}{c}{c$-$o}&
\multicolumn{1}{c}{Design-}&
\multicolumn{1}{c}{J}&
\multicolumn{1}{c}{Energy}&
\multicolumn{1}{c}{c$-$o}&
 \\
\multicolumn{1}{c}{ation}&
\multicolumn{1}{c}{}&
\multicolumn{1}{c}{cm$^{-1}$}&
\multicolumn{1}{c}{cm$^{-1}$}&
\multicolumn{1}{c}{ation}&
\multicolumn{1}{c}{}&
\multicolumn{1}{c}{cm$^{-1}$}&
\multicolumn{1}{c}{cm$^{-1}$}&
\multicolumn{1}{c}{ation}&
\multicolumn{1}{c}{}&
\multicolumn{1}{c}{cm$^{-1}$}&
\multicolumn{1}{c}{cm$^{-1}$}
\\
\noalign{\smallskip}\hline
4[7]   & 15/2 & 124421.468 &$+$12.238 & \\
       & 13/2 & 124436.436 &$+$36.895 & \\
\\
4[6]   &13/2  & 124400.107 &$+$4.567 &  3[6]& 13/2 & 124661.274 &$+$15.827 & \\
       &11/2  & 124402.557 &$-$3.593 &     & 11/2 & 124656.535 &$+$7.092 & \\
\\
4[5]   &11/2  & 124388.840 &$+$3.174 & 3[5]& 11/2 & 124626.900 &$+$3.179  &2[5] & 11/2& 124803.873&$+$20.054\\
       &9/2   & 124385.706 &$+$2.938 &     &  9/2 & 124636.116 &$+$3.120  &     & 9/2 & 124809.727&$+$15.721\\
\\
4[4]   &9/2   & 124401.939 &$+$4.674 & 3[4]&  9/2 & 124623.120 &$+$3.085  &2[4] & 9/2 & 124793.905 &$+$12.624     \\
       &7/2   & 124385.010 &$+$0.698 &     &  7/2 & 124620.914 &$+$7.289  &     & 7/2 & 124783.748 &$+$15.272\\
\\
4[3]   &7/2   & 124416.110&$+$13.187  & 3[3] & 7/2 & 124641.989 &$+$9.092  &2[3] & 7/2 &(124814.025)& \\
       &5/2   & 124403.474 &$+$1.243 &      & 5/2 & 124653.022 &$-$8.651  &     & 5/2 &(124808.178) \\
\\
4[2]   &5/2   & 124434.563 &$+$23.142  &   3[2] & 5/2 &(124670.316)&   & 2[2] & 5/2 &(124835.676) & \\
       &3/2   & 124460.410  &$-$11.802 &        & 3/2 &(124678.325) &   &       & 3/2 &(124833.418) & \\   
\\
4[1]   &3/2   & (124487.989)&  &  3[1] &3/2 &(124697.077) &  &  2[1] & 3/2 &(124876.972)&    \\
       &1/2   & (124484.721)&  &       &1/2 &(124708.453) &   &     & 1/2 &(124874.375)&      \\
\\
       &      &              & & 3[0] &1/2 &  124731.762 &$-$4.875                   \\
\hline
\noalign{\smallskip}
\end{tabular}
\end{flushleft}
\end{table*}

\begin{table*} 
\begin{flushleft}
\caption {\ion{Fe}{ii} energy levels for the  3d$^{6}$\,($^{3}$G)4f subconfiguration. Energies
between parentheses are predicted values.} 
\begin{tabular}{crcrcrcrcrcrrrr}
\noalign{\smallskip}\hline
\multicolumn{1}{c}{Design-}&
\multicolumn{1}{c}{J}&
\multicolumn{1}{c}{Energy}&
\multicolumn{1}{c}{c$-$o}&
\multicolumn{1}{c}{Design-}&
\multicolumn{1}{c}{J}&
\multicolumn{1}{c}{Energy}&
\multicolumn{1}{c}{c$-$o}&
\multicolumn{1}{c}{Design-}&
\multicolumn{1}{c}{J}&
\multicolumn{1}{c}{Energy}&
\multicolumn{1}{c}{c$-$o}&
\\
\multicolumn{1}{c}{ation}&
\multicolumn{1}{c}{}&
\multicolumn{1}{c}{cm$^{-1}$}&
\multicolumn{1}{c}{cm$^{-1}$}&
\multicolumn{1}{c}{ation}&
\multicolumn{1}{c}{}&
\multicolumn{1}{c}{cm$^{-1}$}&
\multicolumn{1}{c}{cm$^{-1}$}&
\multicolumn{1}{c}{ation}&
\multicolumn{1}{c}{}&
\multicolumn{1}{c}{cm$^{-1}$}&
\multicolumn{1}{c}{cm$^{-1}$}
\\
\noalign{\smallskip}\hline
5[8]   & 17/2 & 127507.241& $-$5.657\\
       & 15/2 & 127524.1227&$+$14.501\\
\\
5[7]   & 15/2 & 127484.653 &$-$1.445  &4[7]& 15/2 & 127892.981&$+$4.313\\
       & 13/2 & 127515.235 &$+$2.816  &    & 13/2 & 127895.260&$+$3.367\\
\\
5[6]   &13/2  & 127489.429 &$-$4.823  &4[6]& 13/2 & 127875.000 &$+$2.236 & 3[6] &13/2 & 128110.214&$-$2.182\\
       &11/2  & 127489.977 &$-$0.294  &    & 11/2 & 127880.436 &$+$1.216 &      &11/2 &(128076.012)& \\
\\
5[5]   &11/2  &127482.748&$+$3.147 & 4[5]& 11/2 & 127869.158 &$+$0.993 & 3[5] & 11/2& 128071.171&$-$10.517\\
       &9/2   & (127484.561)& &      &  9/2 & 127855.952 &$-$16.898  &     & 9/2 & 128055.658&$-$16.898\\
\\
5[4]   &9/2   & 127485.362 &$-$15.194 & 4[4]&  9/2 & 127869.892 &$-$4.920 & 3[4] & 9/2 & 128062.710&$-$15.669\\
       &7/2   & 127485.699&$+$9.404  &    &  7/2 & (127871.098)& &       & 7/2 & 128066.823&$-$22.228\\
\\
5[3]   &7/2   &(127476.624)&   & 4[3] & 7/2 &(127877.776) &  & 3[3] & 7/2 & (128047.849)\\
       &5/2   &127510.913 &$+$9.552   &    & 5/2 &   127874.745&$+$5.549 &      & 5/2 & 128063.103&$-$8.192\\
\\
5[2]   &5/2   &(127499.343) & & 4[2] & 5/2 &(127868.807)&  &3[2] & 5/2 & 128089.313&$+$10.032\\
       &3/2   &127487.681 &$-$0.341 &    & 3/2 &(127895.930) & &      & 3/2 & (128069.044) &\\
\\
       &      &           &   &4[1] &3/2 &(127876.787) & &  3[1] & 3/2 &(128099.051) &\\
       &      &           &   &     &1/2 &(127898.510) & &      & 1/2 & (128099.237) &\\
\\
       &      &            &  &     &    &             &   &3[0] &1/2 &(128161.312) & \\

\hline
\noalign{\smallskip}
\end{tabular}
\end{flushleft}
\end{table*}

\section{The new \ion{Fe}{ii} lines}

The new \ion{Fe}{ii} lines in the 3800-8000\,\AA\ region,
produced by transitions to the \ion{Fe}{ii} subconfigurations 
($^{3}$P)4f, ($^{3}$H)4f, ($^{3}$F)4f, and ($^{3}$G)4f,
are shown in Tables\,6$-$9, respectively. Only lines
with $\log\,gf$\,$\ge$\,$-$1.50 are listed, because lines
with lower $\log\,gf$ values are not
observable in this wavelength region of HR\,6000.
 The new \ion{Fe}{ii} lines are mostly concentrated
in the 5100-5400\,\AA\ interval.
The upper energy levels (cols.\,1$-$4) were derived as described in Sect.\,3, 
the lower energy levels (cols.\,5$-$6) are those described in Sect.\,4, the 
calculated wavelength (col.\,7) is the Ritz wavelength in air, the $\log\,gf$ 
values (col.\,8) were computed by Kurucz, the observed wavelengths
(col.\,9) are the wavelengths of lines well observable in the HR\,6000 spectrum. 
Most of them were listed as unidentified lines in 
Castelli \& Hubrig (2007)\footnote{http://wwwuser.oat.ts.astro.it/castelli/hr6000/unidentified.txt}. 
In the last column, comments derived from the comparison of the observed and computed spectra are 
added for most lines.
 In a few cases, 
both computed and observed stellar lines correspond to lines measured by Johansson
in laboratory  works (Johansson 1978; 
 Castelli, Johansson, \& Hubrig 2008). The notes ``J78'' and ``lab''
are added for these lines.  When lines are computed weaker than the observed ones
the disagreement can be due either to a too low $\log\,gf$ value or
to some unknown component which increases the line intensity.
When lines are computed much stronger than the observed ones, some problem with the
energy levels or/and $\log\,gf$ computations is very probably present. 
When we observed a very good agreement between the observed and computed
lines, either isolated or blends, we added the note ``good agreement''.

\begin{table}
\begin{flushleft}
\caption {\ion{Fe}{ii} new levels from  3d$^{6}$4d, 3d$^{6}$5d, and 3d$^{6}$6d configurations.} 
\begin{tabular}{llrlrlrr}
\noalign{\smallskip}\hline
\multicolumn{2}{c}{Designation}&
\multicolumn{1}{c}{J}&
\multicolumn{1}{c}{Energy}&
\multicolumn{1}{c}{c$-$o}
\\
\multicolumn{2}{c}{}&
\multicolumn{1}{c}{}&
\multicolumn{1}{c}{cm$^{-1}$}&
\multicolumn{1}{c}{cm$^{-1}$}
\\
\noalign{\smallskip}\hline
3d$^{6}$($^{3}$P)4d& $^{2}$F   &7/2  & 103191.917 & $+$27.014 \\
3d$^{6}$($^{3}$P)4d& $^{2}$D   &5/2  & 103597.402 & $-$5.701\\
3d$^{6}$($^{3}$F)4d& $^{2}$F   &7/2  & 105775.491 & $-$42.697\\
\\
3d$^{6}$($^{3}$H)5d& $^{4}$H   &13/2  & 124208.725 & $+$47.495\\
3d$^{6}$($^{3}$H)5d& $^{4}$G   &11/2  & 124251.805 & $+$44.041\\
3d$^{6}$($^{3}$H)5d& $^{4}$K   &15/2  & 124297.017 & $-$5.220\\
3d$^{6}$($^{3}$H)5d& $^{4}$I   &15/2  & 124357.304 & $+$12.292\\
3d$^{6}$($^{3}$H)5d& $^{4}$K   &13/2  & 124415.353 & $-$14.256\\
3d$^{6}$($^{3}$H)5d& $^{2}$I   &11/2  & 124976.008 & $-$38.096\\
3d$^{6}$($^{3}$F)5d& $^{4}$H   &13/2  & 125732.991 & $+$9.243\\
\\
3d$^{6}$($^{5}$D)6d& $^{6}$D   &5/2  & 113934.466 & $-$58.836\\
3d$^{6}$($^{5}$D)6d& $^{4}$D   &7/2  & 114009.934 & $-$3.477\\
3d$^{6}$($^{5}$D)6d& $^{6}$G   &7/2  & 114428.399 & $+$51.787\\
3d$^{6}$($^{5}$D)6d& $^{6}$G   &5/2  & 114619.007 & $+$22.415\\
\hline
\noalign{\smallskip}
\end{tabular}
\end{flushleft}
\end{table}

Figure\,1 shows the \ion{Fe}{ii} spectrum in the 5185-5196\,\AA\
interval,  computed before and after the determination of the new 
energy levels.
Figure\,2 compares the observed spectrum of HR\,6000 with the
synthetic spectrum computed with the line list including the new \ion{Fe}{ii}
lines.
When the two figures are considered together, the improvement in the
comparison between the observed and computed spectra is evident.

\begin{figure*}
\centering
\resizebox{5.00in}{!}{\rotatebox{90}{\includegraphics[0,100][500,700]
{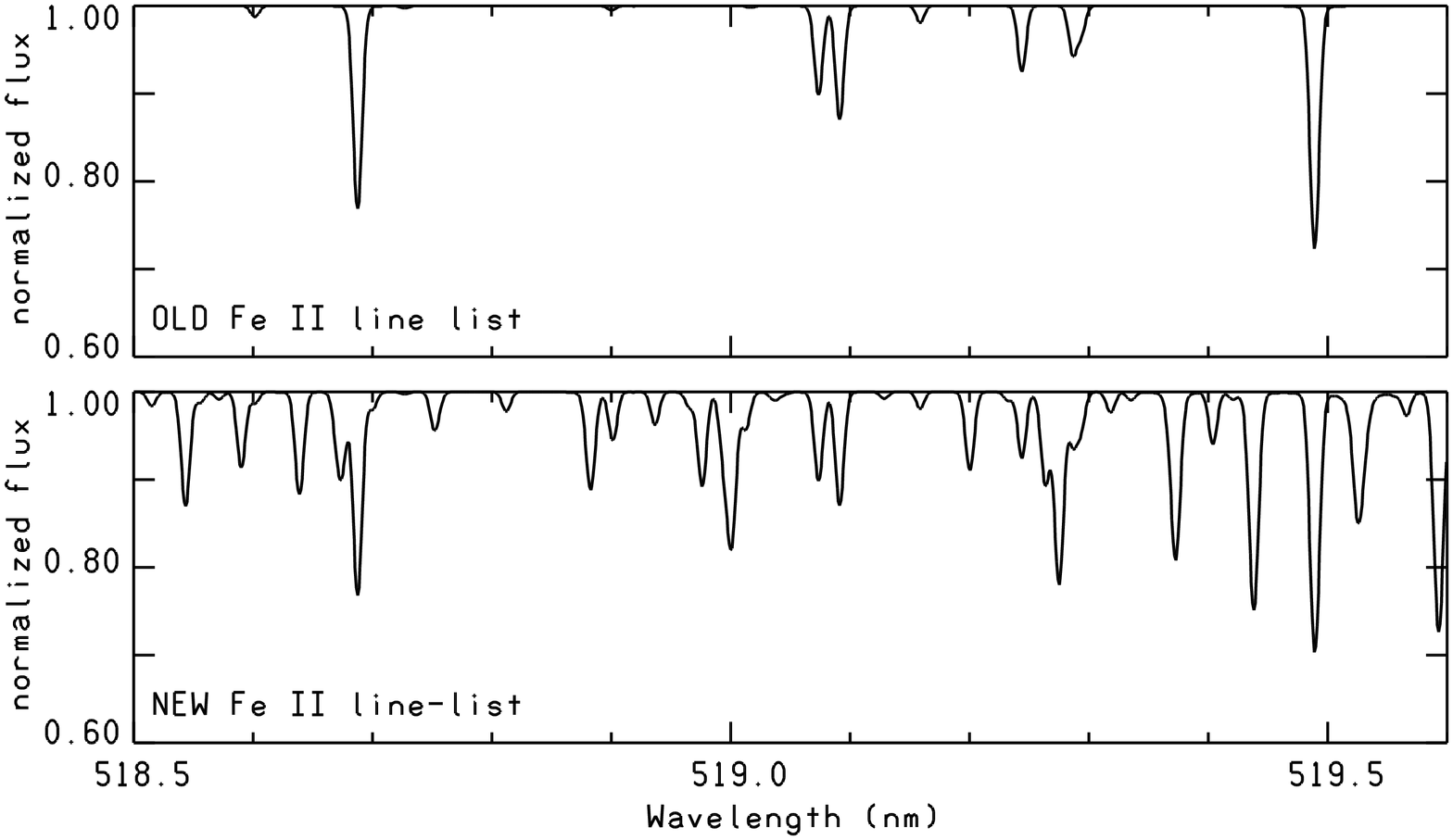}}}
\vskip -0.2cm
\caption{Upper panel shows the \ion{Fe}{ii} synthetic spectrum for the
parameters of HR\,6000 (\teff=13450\,K, \logg=4.3, v{\it sini}=1.5\,km$^{-1}$, 
[Fe/H]]=+0.9) computed with the line list availble before this work.
The lower panel is the same, but with the new \ion{Fe}{ii} lines added in the line list.}

\centering
\resizebox{5.00in}{!}{\rotatebox{90}{\includegraphics[0,100][550,700]
{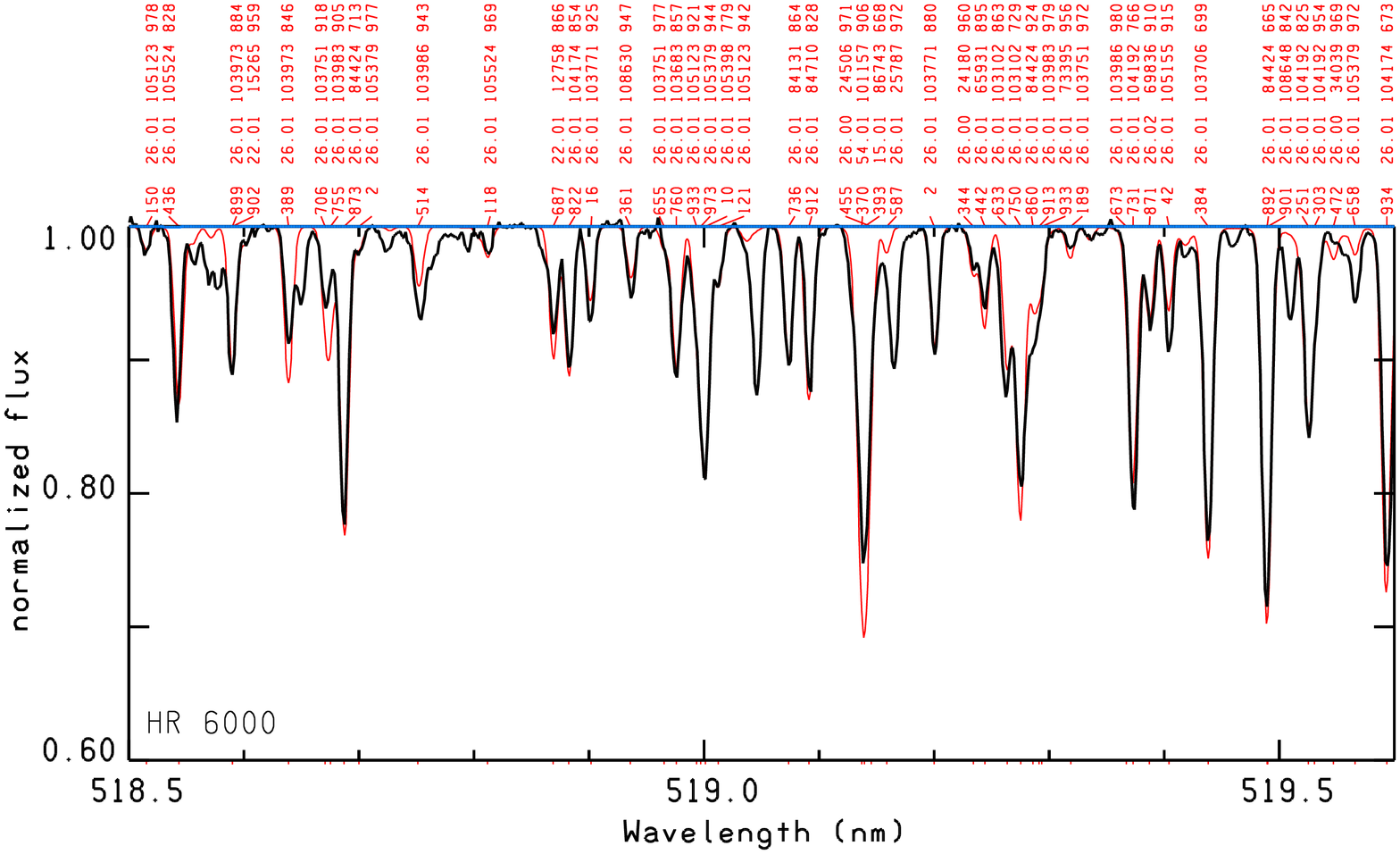}}}
\vskip -1.0cm
\caption{Comparison of the UVES spectrum of HR\,6000 (black line) with a synthetic
spectrum (red line) computed with a line list including the new \ion{Fe}{ii} lines.
The line identification can be decoded as follows: for the first line, 150 last 3 
digits of wavelength 518.5150\,nm; 26 atomic number of iron; .01 charge/100,
i.e. 26.01 identifies the line as \ion{Fe}{ii}; 105123 is the energy of the lower level in cm$^{-1}$;
970 is the residual central intensity in per mil.
}

\end{figure*}

\section{Conclusions}

Computed atomic data and
stellar spectra observed at high resolution 
and high signal-to-noise ratio of the iron$-$overabundant, 
slow$-$rotating star HR\,6000 were used to extend laboratory studies on 
\ion{Fe}{ii} energy levels and line transitions.
We identified as \ion{Fe}{ii} about 500 unidentified spectral lines
in the 3800$-$8000\,\AA\ region. A few of these lines were already
identified as iron from laboratory analyses (Johansson 2007, private communication), but 
they were never classified. Because numerous other new lines are 
components of blends they contribute to improve the agreement between 
observed and  computed spectra. On the other hand, there is a small number
of new lines which are not observed in the spectrum. We believe
that they are  due  to computational problems  related with the
mixing of the even energy levels rather than  to incorrect energy values
for the new 4f odd levels.
 
In spite of the large number of the new identified lines,
several medium-strong lines and a conspicuous number of
weak lines remain still unidentified in the spectral region
we analyzed. If we examine the list of the \ion{Fe}{ii}
lines which correspond to transitions from predicted energy levels, 
we can
count about 4600 lines with $\log\,gf$\,$\ge$\,$-$1.0, where about 400 of them 
have $\log\,gf$\,$\ge$\,0.0. Because the transitions producing these lines
occur between high-excitation energy levels that are not strongly populated,
 most of the lines are weak
in a star like HR\,6000. 
This large number of weak predicted lines 
could explain the spectrum of HR\,6000 longward of about 5800\,\AA.
The spectrum looks like it is affected
by a noise larger than that due to the instrumental 
effects. Castelli \& Hubrig (2007) explained this ``noise'' with
 the presence of a T-Tauri star affecting
the HR\,6000 spectrum. After this study, we prefer to state that
the spectrum shows the presence of  numerous weak \ion{Fe}{ii} lines 
from high-excitation levels, probably 4d, 5d, 6d $-$ 4f, 5f, 6f transitions,
which still have to be identified.
The hypothesis of the presence of the T-Tauri star affecting
the HR\,6000 spectrum  is an example of an incorrect conclusion that 
can be drawn owing to the use of incomplete line lists. 
We will extend this study of the \ion{Fe}{ii} spectrum to
the near infrared region in the near future using CRIRES (CRyogenic high-resolution
InfraRed Echelle Spectrograph) observations of HR\,6000 and 46\,Aql.
The  observations are scheduled
in summer 2010 (ESO proposal 41380, P.\,I.~~S.\, Hubrig).

\Online

\begin{table*}[H] 
\begin{flushleft}
\caption{\ion{Fe}{ii} lines in the 3800-8000\,\AA\ region  
with $\log\,gf$\,$\ge$\,$-$1.5  and 3d$^{6}$($^{3}$P)4f 
energy levels as upper levels } 
\begin{tabular}{llllllllllll}
\noalign{\smallskip}\hline
\multicolumn{4}{c}{Upper level}&
\multicolumn{2}{c}{Lower level}&
\multicolumn{1}{c}{$\lambda$(calc)}&
\multicolumn{1}{c}{$log\,gf$}&
\multicolumn{1}{c}{$\lambda$(obs)}&
\multicolumn{1}{c}{Notes}
\\
\noalign{\smallskip}\hline
\multicolumn{1}{c}{cm$^{-1}$}&
\multicolumn{1}{c}{} &
\multicolumn{1}{c}{} &
\multicolumn{1}{c}{J} &
\multicolumn{1}{c}{cm$^{-1}$}&
\multicolumn{1}{c}{} &
\multicolumn{1}{c}{\AA} &
\multicolumn{1}{c}{KUR} &
\multicolumn{1}{c}{\AA} 
\\
\noalign{\smallskip}\hline
 122351.810 & ($^{3}$P)4f & 2[5] & 11/2 & 103165.320 & ($^{3}$P)4d $^{4}$F$_{9/2}$ & 5210.550 & $+$0.795& 5210.55 & good agreement \\   
               &          &      &      & 103683.070 & ($^{5}$D)5d $^{4}$F$_{9/2}$ & 5355.059 & $+$0.164& 5355.06 & computed too strong\\  
               &          &      &      & 103771.320 & ($^{3}$H)4d $^{4}$G$_{9/2}$ & 5380.493 & $-$1.047&    & at the noise level\\  
               &          &      &      & 104807.210 & ($^{3}$H)4d $^{2}$G$_{9/2}$ & 5698.178 & $-$0.539&    & blend with a telluric line \\  
               &          &      &      & 104916.550 & ($^{3}$H)4d $^{4}$F$_{9/2}$ & 5733.913 & $-$0.635& 5733.90 & computed too weak\\  
               &          &      &      & 106722.170 & ($^{3}$F)4d $^{4}$F$_{9/2}$ & 6396.332 & $-$0.741& 6396.32 & computed too weak\\
\end{tabular}
\end{flushleft}
\end{table*}

\setcounter{table}{5}
\clearpage

\begin{table*}[H] 
\begin{flushleft}
\caption{\ion{Fe}{ii} lines in the 3800-8000\,\AA\ region  
with $\log\,gf$\,$\ge$\,$-$1.5  and 3d$^{6}$($^{3}$P)4f 
energy levels as upper levels } 
 \begin{tabular}{lllllllllll}
\noalign{\smallskip}\hline
\multicolumn{4}{c}{Upper level}&
\multicolumn{2}{c}{Lower level}&
\multicolumn{1}{c}{$\lambda$(calc)}&
\multicolumn{1}{c}{$log\,gf$}&
\multicolumn{1}{c}{$\lambda$(obs)}&
\multicolumn{1}{c}{Notes}
\\
\noalign{\smallskip}\hline
\multicolumn{1}{c}{cm$^{-1}$}&
\multicolumn{1}{c}{} &
\multicolumn{1}{c}{} &
\multicolumn{1}{c}{J} &
\multicolumn{1}{c}{cm$^{-1}$}&
\multicolumn{1}{c}{} &
\multicolumn{1}{c}{\AA} &
\multicolumn{1}{c}{KUR} &
\multicolumn{1}{c}{\AA} 
\\
\noalign{\smallskip}\hline
123637.833 & ($^{3}$P)4f & 1[4] & 7/2     & 102802.312 & ($^{5}$D)6s $^{4}$D$_{5/2}$ & 4798.155 & $-$1.297&  &at the continuum level \\
           &             &      &         & 103002.670 & ($^{3}$P)4d $^{4}$D$_{5/2}$ & 4844.743 & $-$0.954&  &computed too strong\\  
           &             &      &         & 103597.402 & ($^{3}$P)4d $^{2}$D$_{5/2}$ & 4988.521 & $-$0.339& 4988.51 & lab \\  
           &             &      &         & 104107.950 & ($^{3}$P)4d $^{4}$F$_{7/2}$ & 5118.932 & $-$0.819& 5118.95 & lab, computed too weak\\  
           &             &      &         & 104120.270 & ($^{5}$D)5d $^{6}$P$_{5/2}$ & 5122.163 & $-$1.282&         & \\  
           &             &      &         & 104481.590 & ($^{3}$H)4d $^{2}$F$_{7/2}$ & 5218.777 & $-$0.644&         &blend\\
\end{tabular}
\end{flushleft}
\end{table*}

\begin{table*}[H] 
\begin{flushleft}
\caption{\ion{Fe}{ii} lines in the 3800-8000\,\AA\ region with 
$\log\,gf$\,$\ge$\,$-$1.5 and  3d$^{6}$($^{3}$H)4f 
energy level as upper levels } 
\begin{tabular}{lllllllllll}
\noalign{\smallskip}\hline
\multicolumn{4}{c}{Upper level}&
\multicolumn{2}{c}{Lower level}&
\multicolumn{1}{c}{$\lambda$(calc)}&
\multicolumn{1}{c}{$log\,gf$}&
\multicolumn{1}{c}{$\lambda$(obs)}&
\multicolumn{1}{c}{Notes}
\\
\noalign{\smallskip}\hline
\multicolumn{1}{c}{cm$^{-1}$}&
\multicolumn{1}{c}{} &
\multicolumn{1}{c}{} &
\multicolumn{1}{c}{J} &
\multicolumn{1}{c}{cm$^{-1}$}&
\multicolumn{1}{c}{} &
\multicolumn{1}{c}{\AA} &
\multicolumn{1}{c}{KUR} &
\multicolumn{1}{c}{\AA} 
\\
\noalign{\smallskip}\hline
122954.180 & ($^{3}$H)4f & 6[9] & 19/2 & 103644.800 & ($^{3}$H)4d $^{4}$K$_{17/2}$ & 5177.388 & $+$1.169   & 5177.394 & J78, lab, good agreement\\
\\
122952.730 & ($^{3}$H)4f &6[9]  & 17/2    & 103644.800 & ($^{3}$H)4d $^{4}$K$_{17/2}$ & 5177.777 & $-$0.930   &          & blend\\
           &             &      &         & 103706.530 & ($^{3}$H)4d $^{4}$K$_{15/2}$ & 5194.384 & $+$0.798   & 5194.387 & lab, good agreement\\
           &             &      &         & 103878.370 & ($^{3}$H)4d $^{4}$I$_{15/2}$ & 5241.181 & $+$0.558   & 5241.183 & J78, lab, good agreement\\
           &             &      &         & 104119.710 & ($^{3}$H)4d $^{2}$K$_{15/2}$ & 5308.346 & $+$0.518   & 5308.350 & J78,lab, good agreement\\
\end{tabular}
\end{flushleft}
\end{table*}

\setcounter{table}{6}
\clearpage

\begin{table*}[H] 
\begin{flushleft}
\caption{\ion{Fe}{ii} lines in the 3800-8000\,\AA\ region with 
$\log\,gf$\,$\ge$\,$-$1.5 and  3d$^{6}$($^{3}$H)4f energy 
level as upper levels }  
\begin{tabular}{llllllllllll}
\noalign{\smallskip}\hline
\multicolumn{4}{c}{Upper level}&
\multicolumn{2}{c}{Lower level}&
\multicolumn{1}{c}{$\lambda$(calc)}&
\multicolumn{1}{c}{$log\,gf$}&
\multicolumn{1}{c}{$\lambda$(obs)}&
\multicolumn{1}{c}{Notes}
\\
\noalign{\smallskip}\hline
\multicolumn{1}{c}{cm$^{-1}$}&
\multicolumn{1}{c}{} &
\multicolumn{1}{c}{} &
\multicolumn{1}{c}{J} &
\multicolumn{1}{c}{cm$^{-1}$}&
\multicolumn{1}{c}{} &
\multicolumn{1}{c}{\AA} &
\multicolumn{1}{c}{KUR} &
\multicolumn{1}{c}{\AA} 
\\
\noalign{\smallskip}\hline
122990.620 & cont.       &      &         & 105063.550 & ($^{3}$F)4d $^{4}$G$_{11/2}$ & 5576.608 & $-$0.487& 5576.60& computed too strong\\
           &             &      &         & 105763.270 & ($^{3}$F)4d $^{2}$H$_{11/2}$ & 5803.114 & $-$0.380& 5803.12& computed too weak\\
           &             &      &         & 106045.690 & ($^{3}$H)4d $^{2}$H$_{11/2}$ & 5899.835 & $+$0.277& 5899.82 & good agreement\\
           &             &      &         & 108630.429 & ($^{1}$I)5s e$^{2}$I$_{11/2}$ & 6961.775 & $-$1.168&  &at the continuum level\\
           &             &      &         & 109049.600 & ($^{3}$G)4d $^{2}$I$_{13/2}$ & 7171.100 & $-$1.477& & at the continuum level\\
           &             &      &         & 109389.880 & ($^{3}$G)4d $^{2}$I$_{11/2}$ & 7350.516 & $-$1.297& 7350.49 ? & computed too weak ? \\
\end{tabular}
\end{flushleft}
\end{table*}

\setcounter{table}{6}
\clearpage

\begin{table*}[H] 
\begin{flushleft}
\caption{\ion{Fe}{ii} lines in the 3800-8000\,\AA\ region with 
$\log\,gf$\,$\ge$\,$-$1.5 and  3d$^{6}$($^{3}$H)4f energy 
levels as upper levels }  
\begin{tabular}{llllllllllll}
\noalign{\smallskip}\hline
\multicolumn{4}{c}{Upper level}&
\multicolumn{2}{c}{Lower level}&
\multicolumn{1}{c}{$\lambda$(calc)}&
\multicolumn{1}{c}{$log\,gf$}&
\multicolumn{1}{c}{$\lambda$(obs)}&
\multicolumn{1}{c}{Notes}
\\
\noalign{\smallskip}\hline
\multicolumn{1}{c}{cm$^{-1}$}&
\multicolumn{1}{c}{} &
\multicolumn{1}{c}{} &
\multicolumn{1}{c}{J} &
\multicolumn{1}{c}{cm$^{-1}$}&
\multicolumn{1}{c}{} &
\multicolumn{1}{c}{\AA} &
\multicolumn{1}{c}{KUR} &
\multicolumn{1}{c}{\AA} 
\\
\noalign{\smallskip}\hline
122988.215 & ($^{3}$H)4f & 6[4] & 9/2  & 103165.320 & ($^{3}$P)4d $^{4}$F$_{9/2}$ & 5043.266 & $-$0.030& \\
           &             &      &      & 103600.430 & ($^{3}$H)4d $^{4}$G$_{11/2}$ & 5156.450 & $+$0.529& 5156.45 & lab \\
           &             &      &      & 103683.070 & ($^{3}$H)4d $^{4}$F$_{9/2}$  & 5178.524 & $-$0.018& 5178.53 & lab \\
           &             &      &      & 103751.660 & ($^{3}$H)4d $^{4}$H$_{11/2}$ & 5196.989 & $-$0.773&  \\
           &             &      &      & 103771.320 & ($^{3}$H)4d $^{4}$G$_{9/2}$  & 5202.306 & $-$0.787& \\
           &             &      &      & 104765.450 & ($^{3}$H)4d $^{2}$I$_{11/2}$ & 5486.117 & $-$1.286 & \\
\end{tabular}
\end{flushleft}
\end{table*}

\setcounter{table}{6}
\clearpage

\begin{table*}[H] 
\begin{flushleft}
\caption{\ion{Fe}{ii} lines in the 3800-8000\,\AA\ region with 
$\log\,gf$\,$\ge$\,$-$1.5 and  3d$^{6}$($^{3}$H)4f energy 
levels as upper levels } 
\begin{tabular}{llllllllllll}
\noalign{\smallskip}\hline
\multicolumn{4}{c}{Upper level}&
\multicolumn{2}{c}{Lower level}&
\multicolumn{1}{c}{$\lambda$(calc)}&
\multicolumn{1}{c}{$log\,gf$}&
\multicolumn{1}{c}{$\lambda$(obs)}&
\multicolumn{1}{c}{Notes}
\\
\noalign{\smallskip}\hline
\multicolumn{1}{c}{cm$^{-1}$}&
\multicolumn{1}{c}{} &
\multicolumn{1}{c}{} &
\multicolumn{1}{c}{J} &
\multicolumn{1}{c}{cm$^{-1}$}&
\multicolumn{1}{c}{} &
\multicolumn{1}{c}{\AA} &
\multicolumn{1}{c}{KUR} &
\multicolumn{1}{c}{\AA} 
\\
\noalign{\smallskip}\hline
123238.440 & ($^{3}$H)4f& 5[7] & 15/2  & 103617.580 & ($^{3}$H)4d $^{4}$H$_{13/2}$ &  5095.196 & $-$0.836 & 5095.19\\
           &            &      &         & 103706.530 & ($^{3}$H)4d $^{4}$K$_{15/2}$ &  5118.401 & $-$0.254 & 5118.40 & lab\\
           &            &      &         & 103832.050 & ($^{3}$H)4d $^{4}$K$_{13/2}$ &  5151.507 & $-$0.716 &         & blend\\
           &            &      &         & 103878.370 & ($^{3}$H)4d $^{4}$I$_{15/2}$ &  5163.831 & $-$0.599 & 5163.82 & lab\\
           &            &      &         & 104064.670 & ($^{3}$H)4d $^{4}$I$_{13/2}$ &  5214.007 & $+$0.873 & 5214.99 & blend\\
           &            &      &         & 104119.710 & ($^{3}$H)4d $^{2}$K$_{15/2}$ &  5229.017 & $-$0.045 & 5229.030 & lab\\
\end{tabular}
\end{flushleft}
\end{table*}

\setcounter{table}{6}
\clearpage

\begin{table*}[H] 
\begin{flushleft}
\caption{\ion{Fe}{ii} lines in the 3800-8000\,\AA\ region with 
$\log\,gf$\,$\ge$\,$-$1.5 and  3d$^{6}$($^{3}$H)4f 
energy levels as upper levels }  
\begin{tabular}{llllllllllll}
\noalign{\smallskip}\hline
\multicolumn{4}{c}{Upper level}&
\multicolumn{2}{c}{Lower level}&
\multicolumn{1}{c}{$\lambda$(calc)}&
\multicolumn{1}{c}{$log\,gf$}&
\multicolumn{1}{c}{$\lambda$(obs)}&
\multicolumn{1}{c}{Notes}
\\
\noalign{\smallskip}\hline
\multicolumn{1}{c}{cm$^{-1}$}&
\multicolumn{1}{c}{} &
\multicolumn{1}{c}{} &
\multicolumn{1}{c}{J} &
\multicolumn{1}{c}{cm$^{-1}$}&
\multicolumn{1}{c}{} &
\multicolumn{1}{c}{\AA} &
\multicolumn{1}{c}{KUR} &
\multicolumn{1}{c}{\AA} 
\\
\noalign{\smallskip}\hline
123270.340 & cont.       &      &        & 104807.210 & ($^{3}$H)4d $^{2}$G$_{9/2}$ & 5414.696 &$-$0.589 & 5414.7 & blend\\
          &             &      &        & 104916.550 & ($^{3}$H)4d $^{4}$F$_{9/2}$ & 5446.953 & $-$0.182& 5446.95 \\
          &             &      &        & 105063.550 & ($^{3}$F)4d $^{4}$G$_{11/2}$ & 5490.931 &$-$1.162 &                  \\
          &             &      &        & 105155.090 & ($^{3}$F)4d $^{4}$G$_{9/2}$ & 5518.678 & $-$0.927 & & wrong,not observed\\
          &             &      &        & 105763.270 & ($^{3}$F)4d $^{2}$H$_{11/2}$ & 5710.394 &$-$0.287 & 5710.40\\
          &             &      &        & 106045.690 & ($^{3}$H)4d $^{2}$H$_{11/2}$ & 5804.025 &$-$0.029  & 5804.02\\
\end{tabular}
\end{flushleft}
\end{table*}

\setcounter{table}{6}
\clearpage

\begin{table*}[H] 
\begin{flushleft}
\caption{\ion{Fe}{ii} lines in the 3800-8000\,\AA\ region with 
$\log\,gf$\,$\ge$\,$-$1.5 and  3d$^{6}$($^{3}$H)4f 
energy level as upper levels }  
\begin{tabular}{llllllllllll}
\noalign{\smallskip}\hline
\multicolumn{4}{c}{Upper level}&
\multicolumn{2}{c}{Lower level}&
\multicolumn{1}{c}{$\lambda$(calc)}&
\multicolumn{1}{c}{$log\,gf$}&
\multicolumn{1}{c}{$\lambda$(obs)}&
\multicolumn{1}{c}{Notes}
\\
\noalign{\smallskip}\hline
\multicolumn{1}{c}{cm$^{-1}$}&
\multicolumn{1}{c}{} &
\multicolumn{1}{c}{} &
\multicolumn{1}{c}{J} &
\multicolumn{1}{c}{cm$^{-1}$}&
\multicolumn{1}{c}{} &
\multicolumn{1}{c}{\AA} &
\multicolumn{1}{c}{KUR} &
\multicolumn{1}{c}{\AA} 
\\
\noalign{\smallskip}\hline
123258.994& cont.           &       &       & 103771.320 & ($^{3}$H)4d $^{4}$G$_{9/2}$  &  5130.020 & $+$0.269 & 5130.0  & lab\\
          &            &       &       & 103874.260 & ($^{3}$H)4d $^{4}$H$_{9/2}$  &  5157.263 & $-$0.663 &         &blend \\
          &            &       &       & 104481.590 & ($^{3}$H)4d $^{2}$F$_{7/2}$  &  5324.070 & $-$0.506 &         & blend\\
          &            &       &       & 104807.210 & ($^{3}$H)4d $^{2}$G$_{9/2}$  &  5418.025 & $-$0.657 & 5418.02 & lab\\
          &            &       &       & 104916.550 & ($^{3}$H)4d $^{4}$F$_{9/2}$  &  5450.323 & $+$0.051 & 5450.30& wrong, computed too strong\\
          &            &       &       & 105063.550 & ($^{3}$F)4d $^{4}$G$_{11/2}$  &  5494.356 & $-$1.301\\
\end{tabular}
\end{flushleft}
\end{table*}

\setcounter{table}{6}
\clearpage

\begin{table*}[H] 
\begin{flushleft}
\caption{\ion{Fe}{ii} lines in the 3800-8000\,\AA\ region with 
$\log\,gf$\,$\ge$\,$-$1.5 and  3d$^{6}$($^{3}$H)4f energy 
levels as upper levels }  
\begin{tabular}{llllllllllll}
\noalign{\smallskip}\hline
\multicolumn{4}{c}{Upper level}&
\multicolumn{2}{c}{Lower level}&
\multicolumn{1}{c}{$\lambda$(calc)}&
\multicolumn{1}{c}{$log\,gf$}&
\multicolumn{1}{c}{$\lambda$(obs)}&
\multicolumn{1}{c}{Notes}
\\
\noalign{\smallskip}\hline
\multicolumn{1}{c}{cm$^{-1}$}&
\multicolumn{1}{c}{} &
\multicolumn{1}{c}{} &
\multicolumn{1}{c}{J} &
\multicolumn{1}{c}{cm$^{-1}$}&
\multicolumn{1}{c}{} &
\multicolumn{1}{c}{\AA} &
\multicolumn{1}{c}{KUR} &
\multicolumn{1}{c}{\AA} 
\\
\noalign{\smallskip}\hline
123235.165  &  cont.          &       &    & 106767.210 & ($^{3}$F)4d $^{4}$F$_{7/2}$ &  6070.719 & $-$0.626 & 6070.71  \\
            &            &       &    & 110167.280 & ($^{3}$G)4d $^{4}$F$_{7/2}$ &  7650.242 & $-$0.970  \\
            &            &       &    & 110570.300 & ($^{3}$G)4d $^{2}$F$_{7/2}$ &  7893.688 & $-$1.448  \\
\\
123211.159  &($^{3}$H)4f &5[2] & 5/2 & 103193.917 & ($^{3}$P)4d $^{2}$F$_{7/2}$ & 4993.801 &$-$0.145 & 4993.80 & computed too strong \\ 
            &            &      &    & 103921.630 & ($^{3}$H)4d $^{4}$G$_{7/2}$ & 5182.716 &$-$1.163 & 5182.707 & good agreement\\
\end{tabular}
\end{flushleft}
\end{table*}

\setcounter{table}{6}

\begin{table*}[H] 
\begin{flushleft}
\caption{\ion{Fe}{ii} lines in the 3800-8000\,\AA\ region with 
$\log\,gf$\,$\ge$\,$-$1.5 and  3d$^{6}$($^{3}$H)4f energy 
levels as upper levels }  
\begin{tabular}{llllllllllll}
\noalign{\smallskip}\hline
\multicolumn{4}{c}{Upper level}&
\multicolumn{2}{c}{Lower level}&
\multicolumn{1}{c}{$\lambda$(calc)}&
\multicolumn{1}{c}{$log\,gf$}&
\multicolumn{1}{c}{$\lambda$(obs)}&
\multicolumn{1}{c}{Notes}
\\
\noalign{\smallskip}\hline
\multicolumn{1}{c}{cm$^{-1}$}&
\multicolumn{1}{c}{} &
\multicolumn{1}{c}{} &
\multicolumn{1}{c}{J} &
\multicolumn{1}{c}{cm$^{-1}$}&
\multicolumn{1}{c}{} &
\multicolumn{1}{c}{\AA} &
\multicolumn{1}{c}{KUR} &
\multicolumn{1}{c}{\AA} 
\\
\noalign{\smallskip}\hline
123427.119 & ($^{3}$H)4f & 4[6] & 11/2 & 103771.320 & ($^{3}$H)4d $^{4}$G$_{9/2}$  & 5086.139 &$-$0.441 & 5086.15\\
           &             &      &      & 103874.260 & ($^{3}$H)4d $^{4}$H$_{9/2}$  & 5112.917 &$-$0.423 &        & blend\\        
           &             &      &      & 103973.780 & ($^{3}$H)4d $^{4}$K$_{11/2}$ & 5139.074 &$+$0.124 & 5139.10\\
           &             &      &      & 104192.480 & ($^{3}$H)4d $^{4}$I$_{9/2}$  & 5197.506 &$+$0.465 & 5197.56 & blend\\
           &             &      &      & 104315.370 & ($^{3}$H)4d $^{2}$K$_{13/2}$ & 5230.927 &$-$1.051 \\        
           &             &      &      & 104622.300 & ($^{3}$H)4d $^{2}$I$_{13/2}$ & 5316.307 &$-$1.253 \\
\end{tabular}
\end{flushleft}
\end{table*}

\setcounter{table}{6}

\begin{table*}[H] 
\begin{flushleft}
\caption{\ion{Fe}{ii} lines in the 3800-8000\,\AA\ region with 
$\log\,gf$\,$\ge$\,$-$1.5 and  3d$^{6}$($^{3}$H)4f energy 
levels as upper levels }  
 \begin{tabular}{llllllllllll}
\noalign{\smallskip}\hline
\multicolumn{4}{c}{Upper level}&
\multicolumn{2}{c}{Lower level}&
\multicolumn{1}{c}{$\lambda$(calc)}&
\multicolumn{1}{c}{$log\,gf$}&
\multicolumn{1}{c}{$\lambda$(obs)}&
\multicolumn{1}{c}{Notes}
\\
\noalign{\smallskip}\hline
\multicolumn{1}{c}{cm$^{-1}$}&
\multicolumn{1}{c}{} &
\multicolumn{1}{c}{} &
\multicolumn{1}{c}{J} &
\multicolumn{1}{c}{cm$^{-1}$}&
\multicolumn{1}{c}{} &
\multicolumn{1}{c}{\AA} &
\multicolumn{1}{c}{KUR} &
\multicolumn{1}{c}{\AA} 
\\
\noalign{\smallskip}\hline
\hline
123460.690& cont.              &      &      & 106018.640 & ($^{3}$F)4d $^{2}$H$_{9/2}$  & 5731.681 &$-$0.446&  & wrong, not observed\\
          &               &      &      & 106097.520 & ($^{3}$H)4d $^{2}$H$_{9/2}$ & 5757.720& $+$0.118 & 5757.72 & J78, computed too low\\
          &               &      &      & 106722.170 & ($^{3}$F)4d $^{4}$F$_{9/2}$ & 5972.589 & $-$0.946 \\        
          &               &      &      & 106767.210 & ($^{3}$F)4d $^{4}$F$_{7/2}$ & 5988.704 & $-$1.212 \\        
          &               &      &      & 106900.370 & ($^{3}$F)4d $^{2}$G$_{7/2}$ & 6036.859 & $-$0.912 \\        
          &               &      &      & 106924.430 & ($^{3}$F)4d $^{2}$G$_{9/2}$ & 6045.643 & $-$0.124& 6045.65\\
\end{tabular}
\end{flushleft}
\end{table*}

\setcounter{table}{6}

\begin{table*}[H] 
\begin{flushleft}
\caption{\ion{Fe}{ii} lines in the 3800-8000\,\AA\ region with 
$\log\,gf$\,$\ge$\,$-$1.5 and  3d$^{6}$($^{3}$H)4f energy 
levels as upper levels }  
\begin{tabular}{llllllllllll}
\noalign{\smallskip}\hline
\multicolumn{4}{c}{Upper level}&
\multicolumn{2}{c}{Lower level}&
\multicolumn{1}{c}{$\lambda$(calc)}&
\multicolumn{1}{c}{$log\,gf$}&
\multicolumn{1}{c}{$\lambda$(obs)}&
\multicolumn{1}{c}{Notes}
\\
\noalign{\smallskip}\hline
\multicolumn{1}{c}{cm$^{-1}$}&
\multicolumn{1}{c}{} &
\multicolumn{1}{c}{} &
\multicolumn{1}{c}{J} &
\multicolumn{1}{c}{cm$^{-1}$}&
\multicolumn{1}{c}{} &
\multicolumn{1}{c}{\AA} &
\multicolumn{1}{c}{KUR} &
\multicolumn{1}{c}{\AA} 
\\
\noalign{\smallskip}\hline
123430.181 & cont.             &      &   & 105379.430 & ($^{3}$F)4d $^{4}$D$_{5/2}$  & 5538.397 &$-$1.442 &  &at the level of the noise\\
           &              &      &   & 105414.180 & ($^{3}$F)4d $^{4}$G$_{5/2}$  & 5549.080 &$-$0.905 &    & blend\\
           &              &      &   & 105630.750 & ($^{5}$D)5d $^{4}$G$_{5/2}$  & 5616.598 &$-$1.451 &    & blend\\
           &              &      &   & 106846.650 & ($^{3}$F)4d $^{4}$F$_{3/2}$  & 6028.409 &$-$1.085 & 6028.40 & at the level of the noise\\
           &              &      &   & 106866.760 & ($^{3}$F)4d $^{4}$F$_{5/2}$  & 6035.729 &$-$1.269 & \\
           &              &      &   & 107407.800 & ($^{3}$F)4d $^{2}$D$_{5/2}$  & 6239.544 &$-$1.446 & \\
           &              &      &   & 110428.280 & ($^{3}$G)4d $^{4}$F$_{5/2}$  & 7689.067 &$-$1.409 & \\
           &              &      &   & 110609.540 & ($^{3}$G)4d $^{4}$F$_{3/2}$  & 7797.776 &$-$1.406 & \\
\hline
\noalign{\smallskip}
\end{tabular}
\end{flushleft}
\end{table*}

\begin{table*}[H] 
\begin{flushleft}
\caption{\ion{Fe}{ii} lines in the 3800-8000\,\AA\ region with $\log\,gf$\,$\ge$\,$-$1.5 and 3d$^{6}$($^{3}$F)4f energy levels as upper levels.} 
\begin{tabular}{llllllllllll}
\noalign{\smallskip}\hline
\multicolumn{4}{c}{Upper level}&
\multicolumn{2}{c}{Lower level}&
\multicolumn{1}{c}{$\lambda$(calc)}&
\multicolumn{1}{c}{$log\,gf$}&
\multicolumn{1}{c}{$\lambda$(obs)}&
\multicolumn{1}{c}{Notes}
\\
\noalign{\smallskip}\hline
\multicolumn{1}{c}{cm$^{-1}$}&
\multicolumn{1}{c}{} &
\multicolumn{1}{c}{} &
\multicolumn{1}{c}{J} &
\multicolumn{1}{c}{cm$^{-1}$}&
\multicolumn{1}{c}{} &
\multicolumn{1}{c}{\AA} &
\multicolumn{1}{c}{K09} &
\multicolumn{1}{c}{\AA} 
\\
\noalign{\smallskip}\hline
124421.468 & ($^{3}$F)4f & 4[7] & 15/2 & 103617.580 & ($^{3}$H)4d $^{4}$H$_{13/2}$ & 4805.451 & $-$0.972&  4805.42&\\  
           &             &      &      & 104064.670 & ($^{3}$H)4d $^{4}$I$_{13/2}$ & 4910.993 & $-$1.090&        & at the continuum level\\  
           &             &      &      & 104119.710 & ($^{3}$H)4d $^{2}$K$_{15/2}$ & 4924.307 & $-$1.174&        & not obs\\  
           &             &      &      & 104622.300 & ($^{3}$H)4d $^{2}$I$_{13/2}$ & 5049.309 & $-$1.258&  5049.3 & very weak\\  
           &             &      &      & 105288.847 & ($^{3}$F)4d $^{4}$H$_{13/2}$ & 5225.221 & $+$0.974&  5225.229 & lab, J78\\ 
\\
\end{tabular}
\end{flushleft}
\end{table*}

\setcounter{table}{7}

\begin{table*}[H] 
\begin{flushleft}
\caption{\ion{Fe}{ii} lines in the 3800-8000\,\AA\ region with $\log\,gf$\,$\ge$\,$-$1.5 and 3d$^{6}$($^{3}$F)4f energy levels as upper levels.} 
\begin{tabular}{lllllllllll}
\noalign{\smallskip}\hline
\multicolumn{4}{c}{Upper level}&
\multicolumn{2}{c}{Lower level}&
\multicolumn{1}{c}{$\lambda$(calc)}&
\multicolumn{1}{c}{$log\,gf$}&
\multicolumn{1}{c}{$\lambda$(obs)}&
\multicolumn{1}{c}{Notes}
\\
\noalign{\smallskip}\hline
\multicolumn{1}{c}{cm$^{-1}$}&
\multicolumn{1}{c}{} &
\multicolumn{1}{c}{} &
\multicolumn{1}{c}{J} &
\multicolumn{1}{c}{cm$^{-1}$}&
\multicolumn{1}{c}{} &
\multicolumn{1}{c}{\AA} &
\multicolumn{1}{c}{K09} &
\multicolumn{1}{c}{\AA} 
\\
\noalign{\smallskip}\hline
124385.706 & ($^{3}$F)4f & 4[5] & 9/2 & 103771.320 & ($^{3}$H)4d $^{4}$G$_{9/2}$    & 4849.626 & $-$1.159 &   &H$_{\beta}$ wing, not obs.  \\
           &             &      &     & 103986.330 & ($^{3}$H)4d $^{4}$H$_{7/2}$    & 4900.742 & $-$1.404&  & at the continuum level  \\
           &             &      &     & 104807.210 & ($^{3}$H)4d $^{2}$G$_{9/2}$    & 5106.222 & $-$0.305&   &   \\
           &             &      &     & 104993.860 & ($^{3}$F)4d $^{4}$D$_{7/2}$    & 5155.371 & $-$0.195&  5155.37 &computed too strong   \\
           &             &      &     & 105063.550 & ($^{3}$F)4d $^{4}$G$_{11/2}$   & 5173.965 & $-$0.955&  5173.98 &computed too weak\\
           &             &      &     & 105123.000 & ($^{3}$H)4d $^{2}$G$_{7/2}$    & 5189.933 & $-$0.112&         &blend\\
\end{tabular}
\end{flushleft}
\end{table*}

\setcounter{table}{7}

\begin{table*}[H] 
\begin{flushleft}
\caption{\ion{Fe}{ii} lines in the 3800-8000\,\AA\ region with $\log\,gf$\,$\ge$\,$-$1.5 and 3d$^{6}$($^{3}$F)4f energy levels as upper levels.} 
\begin{tabular}{lllllllllll}
\noalign{\smallskip}\hline
\multicolumn{4}{c}{Upper level}&
\multicolumn{2}{c}{Lower level}&
\multicolumn{1}{c}{$\lambda$(calc)}&
\multicolumn{1}{c}{$log\,gf$}&
\multicolumn{1}{c}{$\lambda$(obs)}&
\multicolumn{1}{c}{Notes}
\\
\noalign{\smallskip}\hline
\multicolumn{1}{c}{cm$^{-1}$}&
\multicolumn{1}{c}{} &
\multicolumn{1}{c}{} &
\multicolumn{1}{c}{J} &
\multicolumn{1}{c}{cm$^{-1}$}&
\multicolumn{1}{c}{} &
\multicolumn{1}{c}{\AA} &
\multicolumn{1}{c}{K09} &
\multicolumn{1}{c}{\AA} 
\\
\noalign{\smallskip}\hline
124385.010 & cont.       &      &     & 106924.430 & ($^{3}$F)4d $^{2}$G$_{9/2}$    & 5725.598 & $-$0.824&  5725.60 &\\
           &             &      &     & 107407.800 & ($^{3}$F)4d $^{2}$D$_{5/2}$    & 5888.617 & $-$0.044&  5888.61 &\\
           &             &      &     & 110570.300 & ($^{3}$G)4d $^{2}$F$_{7/2}$    & 7236.667 & $-$1.221&         &at the level of the noise\\
\\
124416.110 & ($^{3}$F)4f & 4[3] & 7/2 & 103683.070 & ($^{5}$D)5d $^{4}$F$_{9/2}$    & 4821.172 & $-$1.273&  \\
           &             &      &     & 104481.590 & ($^{3}$H)4d $^{2}$F$_{7/2}$    & 5015.025 & $-$0.607& 5015.02 &  \\
\end{tabular}
\end{flushleft}
\end{table*}

\setcounter{table}{7}

\begin{table*}[H] 
\begin{flushleft}
\caption{\ion{Fe}{ii} lines in the 3800-8000\,\AA\ region with $\log\,gf$\,$\ge$\,$-$1.5 and 3d$^{6}$($^{3}$F)4f energy levels as upper levels.} 
 \begin{tabular}{lllllllllll}
\noalign{\smallskip}\hline
\multicolumn{4}{c}{Upper level}&
\multicolumn{2}{c}{Lower level}&
\multicolumn{1}{c}{$\lambda$(calc)}&
\multicolumn{1}{c}{$log\,gf$}&
\multicolumn{1}{c}{$\lambda$(obs)}&
\multicolumn{1}{c}{Notes}
\\
\noalign{\smallskip}\hline
\multicolumn{1}{c}{cm$^{-1}$}&
\multicolumn{1}{c}{} &
\multicolumn{1}{c}{} &
\multicolumn{1}{c}{J} &
\multicolumn{1}{c}{cm$^{-1}$}&
\multicolumn{1}{c}{} &
\multicolumn{1}{c}{\AA} &
\multicolumn{1}{c}{K09} &
\multicolumn{1}{c}{\AA} 
\\
\noalign{\smallskip}\hline
124460.410& ($^{3}$F)4f & 4[2] & 3/2 & 104120.270 &  ($^{5}$D)5d $^{6}$P$_{5/2}$ & 4915.015 & $-$1.449&      & \\
           &             &      &     & 104189.380 &  ($^{5}$D)5d $^{4}$P$_{3/2}$ & 4931.772 & $-$1.122&      & wrong,not observed\\
           &             &      &     & 105234.060 &  ($^{3}$H)4d $^{4}$F$_{5/2}$ & 5199.747 & $-$1.496&     & \\
           &             &      &     & 105317.440 &  ($^{3}$P)4d $^{2}$P$_{3/2}$ & 5222.396 & $-$0.923&          & blend  \\
           &             &      &     & 105379.430 &  ($^{3}$F)4d $^{4}$D$_{5/2}$ & 5239.362 & $-$1.350&          & blend  \\
           &             &      &     & 105460.230 &  ($^{3}$F)4d $^{4}$D$_{3/2}$ & 5261.644 & $-$0.436&          &wrong, not observed  \\
\end{tabular}
\end{flushleft}
\end{table*}

\setcounter{table}{7}

\begin{table*}[H] 
\begin{flushleft}
\caption{\ion{Fe}{ii} lines in the 3800-8000\,\AA\ region with $\log\,gf$\,$\ge$\,$-$1.5 and 3d$^{6}$($^{3}$F)4f energy levels as upper levels.} 
 \begin{tabular}{lllllllllll}
\noalign{\smallskip}\hline
\multicolumn{4}{c}{Upper level}&
\multicolumn{2}{c}{Lower level}&
\multicolumn{1}{c}{$\lambda$(calc)}&
\multicolumn{1}{c}{$log\,gf$}&
\multicolumn{1}{c}{$\lambda$(obs)}&
\multicolumn{1}{c}{Notes}
\\
\noalign{\smallskip}\hline
\multicolumn{1}{c}{cm$^{-1}$}&
\multicolumn{1}{c}{} &
\multicolumn{1}{c}{} &
\multicolumn{1}{c}{J} &
\multicolumn{1}{c}{cm$^{-1}$}&
\multicolumn{1}{c}{} &
\multicolumn{1}{c}{\AA} &
\multicolumn{1}{c}{K09} &
\multicolumn{1}{c}{\AA} 
\\
\noalign{\smallskip}\hline
124636.116  & cont.     &        &       & 104993.860 & ($^{3}$F)4d $^{4}$D$_{7/2}$ & 5089.646 &$-$0.797 &          &weak \\
            &            &        &       & 105123.000 & ($^{3}$H)4d $^{2}$G$_{7/2}$ & 5123.331 &$-$1.032 &          & \\
           &            &        &       & 105155.090 & ($^{3}$F)4d $^{4}$G$_{9/2}$ & 5131.770 &$-$0.298 &          &blend \\
           &            &        &       & 105211.062 & ($^{5}$D)5d $^{4}$G$_{9/2}$ & 5146.557 &$-$0.622 &          &blend \\
           &            &        &       & 105220.600 & ($^{3}$H)4d $^{4}$F$_{7/2}$ & 5149.085 &$+$0.286 & 5149.1   &lab \\
           &            &        &       & 105291.010 & ($^{3}$F)4d $^{4}$G$_{7/2}$ & 5167.827.&$-$0.884 & 5167.82  &computed too weak \\
\end{tabular}
\end{flushleft}
\end{table*}

\setcounter{table}{7}

\begin{table*}[H] 
\begin{flushleft}
\caption{\ion{Fe}{ii} lines in the 3800-8000\,\AA\ region with $\log\,gf$\,$\ge$\,$-$1.5 and 3d$^{6}$($^{3}$F)4f energy levels as upper levels.} 
\begin{tabular}{lllllllllll}
\noalign{\smallskip}\hline
\multicolumn{4}{c}{Upper level}&
\multicolumn{2}{c}{Lower level}&
\multicolumn{1}{c}{$\lambda$(calc)}&
\multicolumn{1}{c}{$log\,gf$}&
\multicolumn{1}{c}{$\lambda$(obs)}&
\multicolumn{1}{c}{Notes}
\\
\noalign{\smallskip}\hline
\multicolumn{1}{c}{cm$^{-1}$}&
\multicolumn{1}{c}{} &
\multicolumn{1}{c}{} &
\multicolumn{1}{c}{J} &
\multicolumn{1}{c}{cm$^{-1}$}&
\multicolumn{1}{c}{} &
\multicolumn{1}{c}{\AA} &
\multicolumn{1}{c}{K09} &
\multicolumn{1}{c}{\AA} 
\\
\noalign{\smallskip}\hline
124641.989 & ($^{3}$F)4f& 3[3]  &7/2     & 104107.950 & ($^{3}$P)4d $^{4}$F$_{7/2}$ & 4868.603 &$-$1.393 &         & \\
           &            &        &       & 104120.270 & ($^{5}$D)5d $^{6}$P$_{5/2}$ & 4871.525 &$-$1.423 &         & \\
           &            &        &       & 104481.590 & ($^{3}$H)4d $^{2}$F$_{7/2}$ & 4958.835 &$-$1.370 &         &blend \\
           &            &        &       & 105123.000 & ($^{3}$H)4d $^{2}$G$_{7/2}$ & 5121.789 &$-$0.828 &         & \\
           &            &        &       & 105155.090 & ($^{3}$F)4d $^{4}$G$_{9/2}$ & 5130.223 &$-$0.928 &         &blend \\
           &            &        &       & 105211.062 & ($^{5}$D)5d $^{4}$G$_{9/2}$ & 5145.002 &$-$1.290 &         &at the level of the noise \\
\end{tabular}
\end{flushleft}
\end{table*}

\setcounter{table}{7}
\clearpage

\begin{table*}[H] 
\begin{flushleft}
\caption{\ion{Fe}{ii} lines in the 3800-8000\,\AA\ region with $\log\,gf$\,$\ge$\,$-$1.5 and 3d$^{6}$($^{3}$F)4f energy levels as upper levels.}  
\begin{tabular}{lllllllllll}
\noalign{\smallskip}\hline
\multicolumn{4}{c}{Upper level}&
\multicolumn{2}{c}{Lower level}&
\multicolumn{1}{c}{$\lambda$(calc)}&
\multicolumn{1}{c}{$log\,gf$}&
\multicolumn{1}{c}{$\lambda$(obs)}&
\multicolumn{1}{c}{Notes}
\\
\noalign{\smallskip}\hline
\multicolumn{1}{c}{cm$^{-1}$}&
\multicolumn{1}{c}{} &
\multicolumn{1}{c}{} &
\multicolumn{1}{c}{J} &
\multicolumn{1}{c}{cm$^{-1}$}&
\multicolumn{1}{c}{} &
\multicolumn{1}{c}{\AA} &
\multicolumn{1}{c}{K09} &
\multicolumn{1}{c}{\AA} 
\\
\noalign{\smallskip}\hline
124803.873 & ($^{3}$F)4f& 2[5]  &11/2 & 103771.320 & ($^{3}$H)4d $^{4}$G$_{9/2}$ & 4753.206 &$-$1.359 & &  \\
           &            &       &     & 104807.210 & ($^{3}$H)4d $^{2}$G$_{9/2}$ & 4999.441 &$-$1.315 & &  \\
           &            &       &        & 105524.461 & ($^{3}$F)4d $^{4}$H$_{9/2}$ & 5185.437 &$+$0.377 &5185.422 & lab \\
           &            &       &         & 106018.643 & ($^{3}$F)4d $^{2}$H$_{9/2}$ & 5321.852 &$+$0.731 &5321.83 & lab \\
           &            &       &         & 106097.520 & ($^{3}$H)4d $^{2}$H$_{9/2}$ & 5344.292 &$-$1.008 &5344.28 & \\
           &            &       &         & 106924.430 & ($^{3}$F)4d $^{2}$G$_{9/2}$ & 5591.464 &$-$0.173 & & computed too strong \\
\end{tabular}
\end{flushleft}
\end{table*}

\begin{table*}[H] 
\begin{flushleft}
\caption{\ion{Fe}{ii} lines in the 3800-8000\,\AA\ region with  
$\log\,gf$\,$\ge$\,$-$1.5 and 3d$^{6}$($^{3}$G)4f 
energy levels as upper levels.}                          
\begin{tabular}{llllllllllll}
\noalign{\smallskip}\hline
\multicolumn{4}{c}{Upper level}&
\multicolumn{2}{c}{Lower level}&
\multicolumn{1}{c}{$\lambda$(calc)}&
\multicolumn{1}{c}{$log\,gf$}&
\multicolumn{1}{c}{$\lambda$(obs)}&
\multicolumn{1}{c}{Notes}
\\
\noalign{\smallskip}\hline
\multicolumn{1}{c}{cm$^{-1}$}&
\multicolumn{1}{c}{} &
\multicolumn{1}{c}{} &
\multicolumn{1}{c}{J} &
\multicolumn{1}{c}{cm$^{-1}$}&
\multicolumn{1}{c}{} &
\multicolumn{1}{c}{\AA} &
\multicolumn{1}{c}{KUR} &
\multicolumn{1}{c}{\AA} 
\\
\noalign{\smallskip}\hline
127507.241 & ($^{3}$G)4f & 5[8] & 17/2 & 103878.370 & ($^{3}$H)4d $^{4}$I$_{15/2}$ & 4230.919 & $-$1.017& 4230.93    &\\  
           &             &      &         & 108337.860 & ($^{3}$G)4d $^{4}$I$_{15/2}$ & 5215.200 & $+$1.119& 5215.21   &\\
\\
127524.122 & ($^{3}$G)4f & 5[8] & 15/2 & 104064.670 & ($^{3}$H)4d $^{4}$I$_{13/2}$ & 4261.475 & $-$1.477&    &\\  
           &             &      &      & 104622.300 & ($^{3}$H)4d $^{2}$I$_{13/2}$ & 4365.238 & $-$1.210&    &\\  
           &             &      &      & 108133.440 & ($^{3}$G)4d $^{4}$H$_{13/2}$ & 5155.680 & $-$0.971& &\\
\end{tabular}
\end{flushleft}
\end{table*}

\setcounter{table}{8}
\clearpage

\begin{table*}[H] 
\begin{flushleft}
\caption{\ion{Fe}{ii} lines in the 3800-8000\,\AA\ region with  
$\log\,gf$\,$\ge$\,$-$1.5 and 3d$^{6}$($^{3}$G)4f 
energy levels as upper levels.} 
\begin{tabular}{lllllllllll}
\noalign{\smallskip}\hline
\multicolumn{4}{c}{Upper level}&
\multicolumn{2}{c}{Lower level}&
\multicolumn{1}{c}{$\lambda$(calc)}&
\multicolumn{1}{c}{$log\,gf$}&
\multicolumn{1}{c}{$\lambda$(obs)}&
\multicolumn{1}{c}{Notes}
\\
\noalign{\smallskip}\hline
\multicolumn{1}{c}{cm$^{-1}$}&
\multicolumn{1}{c}{} &
\multicolumn{1}{c}{} &
\multicolumn{1}{c}{J} &
\multicolumn{1}{c}{cm$^{-1}$}&
\multicolumn{1}{c}{} &
\multicolumn{1}{c}{\AA} &
\multicolumn{1}{c}{KUR} &
\multicolumn{1}{c}{\AA} 
\\
\noalign{\smallskip}\hline
127482.748 & cont.       &      &       & 108463.910 & ($^{3}$G)4d $^{4}$I$_{13/2}$ & 5256.482 & $-$0.830& 5256.5 &\\  
           &             &      &       & 108648.695 & ($^{1}$I)5s e$^{2}$I$_{13/2}$ & 5308.055 & $-$1.341& &\\  
           &             &      &       & 108775.080 & ($^{3}$G)4d $^{4}$I$_{11/2}$ & 5343.915 & $-$1.043& &\\  
           &             &      &       & 109625.200 & ($^{3}$G)4d $^{2}$G$_{9/2}$ & 5598.319 & $-$0.100&  5598.32& computed too weak\\  
           &             &      &       & 109683.280 & ($^{3}$G)4d $^{2}$H$_{11/2}$ & 5616.586 & $-$0.042& 5616.6 & computed too weak\\  
           &             &      &       & 109811.920 & ($^{3}$G)4d $^{4}$F$_{9/2}$ & 5657.474 & $-$0.662& 5657.50& computed too weak\\
\end{tabular}
\end{flushleft}
\end{table*}

\setcounter{table}{8}
\clearpage

\begin{table*}[H] 
\begin{flushleft}
\caption{\ion{Fe}{ii} lines in the 3800-8000\,\AA\ region with  
$\log\,gf$\,$\ge$\,$-$1.5 and 3d$^{6}$($^{3}$G)4f 
energy levels as upper levels.}                                                    
\begin{tabular}{lllllllllll}
\noalign{\smallskip}\hline
\multicolumn{4}{c}{Upper level}&
\multicolumn{2}{c}{Lower level}&
\multicolumn{1}{c}{$\lambda$(calc)}&
\multicolumn{1}{c}{$log\,gf$}&
\multicolumn{1}{c}{$\lambda$(obs)}&
\multicolumn{1}{c}{Notes}
\\
\noalign{\smallskip}\hline
\multicolumn{1}{c}{cm$^{-1}$}&
\multicolumn{1}{c}{} &
\multicolumn{1}{c}{} &
\multicolumn{1}{c}{J} &
\multicolumn{1}{c}{cm$^{-1}$}&
\multicolumn{1}{c}{} &
\multicolumn{1}{c}{\AA} &
\multicolumn{1}{c}{KUR} &
\multicolumn{1}{c}{\AA} 
\\
\noalign{\smallskip}\hline
127892.981 & ($^{3}$G)4f & 4[7] & 15/2    & 104064.670 & ($^{3}$H)4d $^{4}$I$_{13/2}$ & 4195.506 & $-$1.455& &\\  
           &             &      &         & 104622.300 & ($^{3}$H)4d $^{2}$I$_{13/2}$ & 4296.044 & $-$1.387& &\\  
           &             &      &         & 108133.440 & ($^{3}$G)4d $^{4}$H$_{13/2}$ & 5059.436 & $-$0.484& 5059.42 & lab\\  
           &             &      &         & 108463.910 & ($^{3}$G)4d $^{4}$I$_{13/2}$ & 5145.493 & $-$0.007& 5145.5  &\\  
           &             &      &         & 108648.695 & ($^{1}$I)5s e$^{2}$I$_{13/2}$ & 5194.901 & $+$0.482& &blend\\  
           &             &      &         & 109049.600 & ($^{3}$G)4d $^{2}$I$_{13/2}$  & 5305.427 & $+$0.862& 5305.42 & lab\\
\end{tabular}
\end{flushleft}
\end{table*}

\setcounter{table}{8}

\begin{table*}[H] 
\begin{flushleft}
\caption{\ion{Fe}{ii} lines in the 3800-8000\,\AA\ region with  
$\log\,gf$\,$\ge$\,$-$1.5 and 3d$^{6}$($^{3}$G)4f 
energy levels as upper levels.}                                                                                                            
\begin{tabular}{lllllllllll}
\noalign{\smallskip}\hline
\multicolumn{4}{c}{Upper level}&
\multicolumn{2}{c}{Lower level}&
\multicolumn{1}{c}{$\lambda$(calc)}&
\multicolumn{1}{c}{$log\,gf$}&
\multicolumn{1}{c}{$\lambda$(obs)}&
\multicolumn{1}{c}{Notes}
\\
\noalign{\smallskip}\hline
\multicolumn{1}{c}{cm$^{-1}$}&
\multicolumn{1}{c}{} &
\multicolumn{1}{c}{} &
\multicolumn{1}{c}{J} &
\multicolumn{1}{c}{cm$^{-1}$}&
\multicolumn{1}{c}{} &
\multicolumn{1}{c}{\AA} &
\multicolumn{1}{c}{KUR} &
\multicolumn{1}{c}{\AA} 
\\
\noalign{\smallskip}\hline
127855.952 &  cont.      &      &         & 109683.280 & ($^{3}$G)4d $^{2}$H$_{11/2}$ & 5501.240 & $-$0.659&  & \\  
           &             &      &         & 109811.920 & ($^{3}$G)4d $^{4}$F$_{9/2}$ & 5540.460 & $-$0.431& 5540.47  & \\  
           &             &      &         & 109901.500 & ($^{3}$G)4d $^{2}$G$_{7/2}$ & 5568.103 & $-$0.216& 5568.10 & \\  
           &             &      &         & 110167.280 & ($^{3}$G)4d $^{4}$F$_{7/2}$ & 5651.767 & $-$0.160& 5651.78 & computed too weak \\  
           &             &      &         & 110570.300 & ($^{3}$G)4d $^{2}$F$_{7/2}$ & 5783.541 & $-$0.854&  & \\  
\\
\end{tabular}
\end{flushleft}
\end{table*}

\setcounter{table}{8}

\begin{table*}[H] 
\begin{flushleft}                                                                                                                        
\caption{\ion{Fe}{ii} lines in the 3800-8000\,\AA\ region with  
$\log\,gf$\,$\ge$\,$-$1.5 and 3d$^{6}$($^{3}$G)4f 
energy levels as upper levels.}
\begin{tabular}{lllllllllll}
\noalign{\smallskip}\hline
\multicolumn{4}{c}{Upper level}&
\multicolumn{2}{c}{Lower level}&
\multicolumn{1}{c}{$\lambda$(calc)}&
\multicolumn{1}{c}{$log\,gf$}&
\multicolumn{1}{c}{$\lambda$(obs)}&
\multicolumn{1}{c}{Notes}
\\
\noalign{\smallskip}\hline
\multicolumn{1}{c}{cm$^{-1}$}&
\multicolumn{1}{c}{} &
\multicolumn{1}{c}{} &
\multicolumn{1}{c}{J} &
\multicolumn{1}{c}{cm$^{-1}$}&
\multicolumn{1}{c}{} &
\multicolumn{1}{c}{\AA} &
\multicolumn{1}{c}{KUR} &
\multicolumn{1}{c}{\AA} 
\\
\noalign{\smallskip}\hline
128055.658 & ($^{3}$F)4f& 3[5]   & 9/2   & 106097.520 & ($^{3}$H)4d $^{2}$H$_{9/2}$ & 4552.844 &$-$1.204 &  & no spectrum \\
           &            &        &       & 106767.210 & ($^{3}$F)4d $^{4}$F$_{7/2}$ & 4696.069 &$-$0.812 &  & no spectrum\\
           &            &        &       & 106924.430 & ($^{3}$F)4d $^{2}$G$_{9/2}$ & 4731.009 &$-$1.380 &  & no spectrum\\
           &            &        &       & 108537.610 & ($^{3}$G)4d $^{4}$G$_{7/2}$ & 5122.036 &$+$0.148 & 5122.02  & lab \\
           &            &        &       & 108577.560 & ($^{3}$G)4d $^{4}$H$_{9/2}$ & 5132.541 &$+$0.038 & 5132.55  & lab \\
           &            &        &       & 108709.450 & ($^{3}$G)4d $^{4}$H$_{7/2}$ & 5167.532 &$-$0.521 &  & blend \\
\end{tabular}
\end{flushleft}
\end{table*}


\begin{thebibliography}{}


\bibitem[Adam et al.(1987)]{1987ApJ...312..337A} Adam, J., Baschek, B., 
Johansson, S., Nilsson, A.~E., \& Brage, T.\ 1987, \apj, 312, 337 

\bibitem[Bi{\'e}mont et al.(1997)]{1997PhyS...55..559B} Bi{\'e}mont, E., 
Johansson, S., \& Palmeri, P.\ 1997, \physscr, 55, 559 

\bibitem[Castelli 
\& Hubrig(2007)]{2007A&A...475.1041C} Castelli, F., \& Hubrig, S.\ 2007, \aap, 475, 1041 

\bibitem[Castelli et al.(2008)]{2008JPhCS.130a2003C} Castelli, F., 
Johansson, S., 
\& Hubrig, S.\ 2008, Journal of Physics Conference Series, 130, 012003 


\bibitem[Castelli et 
al.(2009)]{2009A&A...508..401C} Castelli, F., Kurucz, R., \& Hubrig, S.\ 2009, \aap, 508, 401 (Paper I) 

\bibitem[1981]{Cowan1981}
Cowan, R.~D.\ 1981,
The Theory of Atomic Structure and Spectra (Berkeley: Univ. California Press)


\bibitem[Johansson(1978)]{1978PhyS...18..217J} Johansson, S.\ 1978, 
\physscr, 18, 217 

\bibitem[Johansson(2009)]{2009PhST..134a4013J} Johansson, S.\ 2009,  
\physscr, T134, 014013 

\bibitem[Johansson 
\& Baschek(1988)]{1988NIMPB..31..222J} Johansson, S., \& Baschek, B.\ 1988, Nuclear Instruments and Methods in Physics Research B, 31, 222 


\bibitem[Kurucz(2005)]{2005MSAIS...8...14K} Kurucz, R.~L.\ 2005, Memorie 
della Societa' Astronomica Italiana, Supplementi, 8, 14 

\bibitem[Kurucz(2009)]{2009AIPC.1171...43K} Kurucz, R.~L.\ 2009, American 
Institute of Physics Conference Series, 1171, 43 

\bibitem[Rosberg 
\& Johansson(1992)]{1992PhyS...45..590R} Rosberg, M., \& Johansson, S.\ 1992, \physscr, 45, 590 

\bibitem[Sugar 
\& Corliss(1985)]{1985aeli.book.....S} Sugar, J., \& Corliss, C.\ 1985, J. Phys. Chem. Ref. Data, 14, Supp.\,2  


\end{thebibliography}
\end{document}